# Process Design and Economics of Production of *p*-Aminophenol

**Chinmay Ghoroi**
Professor at Department of Chemical Engineering
Indian Institute of Technology - Gandhinagar
Gandhinagar, 382355
chinmayg@iitgn.ac.in

**Jay Shah**
Department of Chemical Engineering
Indian Institute of Technology - Gandhinagar
Gandhinagar, 382355
jay.shah@iitgn.ac.in

**Devanshu Thakar**
Department of Chemical Engineering
Indian Institute of Technology - Gandhinagar
Gandhinagar, 382355
nilesh.thakar@iitgn.ac.in

**Sakshi Baheti**
Department of Chemical Engineering
Indian Institute of Technology - Gandhinagar
Gandhinagar, 382355
sakshi.baheti@iitgn.ac.in

October 29, 2021

## ABSTRACT

Para-Aminophenol is one of the key chemicals required for the synthesis of Paracetamol, an analgesic and antipyretic drug. Data shows a large fraction of India's demand for Para-Aminophenol being met through imports from China. The uncertainty in the India-China relations would affect the supply and price of this "Key Starting Material." This report is a detailed business plan for setting up a plant and producing Para-Aminophenol in India at a competitive price. The plant is simulated in AspenPlus V8 and different Material Balances and Energy Balances calculations are carried out. The plant produces 22.7 kmols Para-Aminophenol per hour with a purity of 99.9 %. Along with the simulation, economic analysis is carried out for this plant to determine the financial parameters like Payback Period and Return on Investment.

# 1. INTRODUCTION

## 1.1 PROPERTIES OF P-AMINOPHENOL

Para-Aminophenol (also known as 4-Aminophenol or para-hydroxyaniline) is a basic amino-phenolic organic compound. It is one of the three isomeric forms of Amino-phenol, which are Ortho-Aminophenol, Meta-Aminophenol and Para-Aminophenol. Its molecular formula is $NH_2C_6H_4OH$ and its structure consists of an Aromatic ring of six Carbon atoms, with the phenol and amino groups attached to the ring at para positions with respect to each other. Its molecular weight is 109 g/mol.

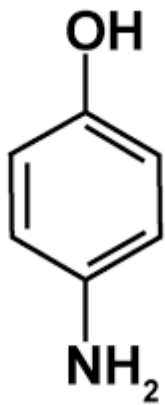

Mostly available in the form of white powder, the other forms of p-aminophenol include light brown powder and white or reddish-yellow crystals. The crystals are either present as orthorhombic pyramidal structures (alpha form, stable) or as needle shaped crystals (beta form, less stable). Under atmospheric conditions, the boiling point of p-aminophenol is 284 ℃ and its melting point is 187.5 ℃.

Para-aminophenol is usually unstable under normal atmospheric conditions as it is sensitive to the presence of air, humidity and light. Hence, its colour changes when exposed to any of these. It acts as a powerful reducing agent, and reacts quickly when exposed to oxygen in an alkaline or basic medium. While p-aminophenol is slightly soluble in cold water and moderately soluble in hot water, it is highly soluble in ethanol, ethyl-methyl ketone and dimethyl sulphoxide and insoluble in Chloroform and Benzene.

Para-aminophenol is a toxic chemical causing the conversion of haemoglobin to methaemoglobin, thereby affecting the normal transport of oxygen via haemoglobin in the human body. It is a poisonous substance which can affect the functioning of kidneys and other organs of the body. It acts as an irritant for skin can and hence its prolonged exposure can result in dermatitis. Moreover, when heated to decompose, it effluiviates vapours of nitric oxides, which are poisonous and hence can result in health hazards.

## 1.2 USES OF PARA-AMINOPHENOL

The main and the most significant use of p-aminophenol is for the manufacturing of Paracetamol, an analgesic and antipyretic drug. In addition to paracetamol, it is a key element in the synthesis of pharmaceutical ingredients and important industrial chemicals like Acebutolol, Ambroxol, Sorafenib and so on.

Para-Aminophenol is widely used in preparation of fabric dyes and hair dyes, and is also used as a developing agent in photography for creating black and white images. It acts as a corrosion inhibitor in paints and as anti-corrosive lubricating agent in 2-cycle engines. It is also used as a wood stain, giving rose-like colour to timber. Para-Aminophenol is one of key ingredients for synthesis of rubber antioxidants. Moreover, it is often used as a reagent for analysing metals like Copper, Magnesium, Vanadium and Gold, compounds like Nitrites and Cyanates, and antioxidants.

## 1.3 MARKET SEGMENTATION AND COMPETITIVE ANALYSIS

Para-Aminophenol is classified into different categories based on its purity. The most commonly available qualities are 98% pure, 99% pure and 99.5% pure. Based on the purity, its cost also varies. Based on its quality and physical state, the price range of para-aminophenol purchased through retailers in India varies from Rs 400/kg to Rs 1000/kg.

As per the data shared by a 2018 report, globally more than 80% of para-aminophenol is used for the synthesis of Paracetamol, whereas nearly 5% of it is used for preparing rubber antioxidants, around 7% is used for dyes, and the remaining is used for other purposes[6]. According to a report by Marketwatch, "The global Para-aminophenol (PAP) market is valued at 442.6 million USD in 2020 is expected to reach 553.1 million USD by the end of 2026, growing at a CAGR of 3.2% during 2021-2026." [5]

Currently, China is the largest producer of p-aminophenol, having a manufacturing capacity of nearly 1.1 lac tonnes per annum (TPA). Globally the major producers of p-aminophenol are Taixing Yangzi (35000 TPA), Liaoning Shixing (40000 TPA), Anuhi Bayi Chemical (60000 TPA) etc. [7]

India is one of the major consumers of p-aminophenol. In India, the demand for p-aminophenol is nearly 40,000 TPA, and most of it (around 28,000 to 30,000 TPA) is met through imports from countries like China. A few local producers of para-aminophenol in India include Bharat Chemicals, Vinati Organics, Aarti Industries and Jay Organics.

In the year 2020, due to the COVID-19 pandemic, the supply of p-aminophenol along with other essential Key Starting Materials (KSMs) and Active Pharmaceutical Ingredients (APIs) from China saw a sudden surge in the prices. If we specifically consider p-aminophenol, its price increased by 100% as compared to the previous year. While in the previous year the price of p-aminophenol imported from China was around $3.5/kg, the following year it increased to around $7.5/kg. Moreover, the constant faceoff between India and China at the border added fuel to fire.

The outburst in the prices due to the pandemic and the ongoing friction with China led the Indian Government introduce a "Production Linked Incentive" (PLI) scheme to reduce our dependence on China and promote the manufacturing of these Key Starting Materials (KSMs) and Active Pharmaceutical Ingredients (APIs) in our country itself. Para-Aminophenol being the main ingredient required for the production of Paracetamol, also comes under the list of KSMs and hence is eligible for the PLI scheme.[9]

## 2. PRELIMINARY PROCESS SYNTHESIS

## 2.1 AVAILABLE PROCESSES

There are many reaction routes available for production of p-aminophenol. Para-Aminophenol can be produced by the reduction of *p*-nitrophenol using iron as a catalyst. *P*-nitrophenol is produced by the phenol nitrosation method which is followed by reduction and acid precipitation. Aniline can also be used to produce *p*-aminophenol through diazotization coupling followed by the reduction with Fe-powder. The drawback of all the above methods is that they involve production of aromatic compounds which are hard to dispose of safely, and some of these processes also involve evolution of polluting gases like oxides of sulphur. Moreover, many of the above processes are multi-step processes, thereby increasing the cost of production. An environmental friendly process involving the reduction of nitrobenzene with hydrogen using Pt-C or Pd-C catalyst is available. We selected this process because it is a single step process and it does not produce any hazardous by-products.

## 2.2 PROCESS OVERVIEW

Conventionally p-aminophenol was manufactured using iron-acid reduction of p-nitrobenzene. Reduction using iron-acid is a multi-step process. The modern method is the catalytic dehydrogenation of nitrobenzene to p-aminophenol (PAP) using a noble metal catalyst in the presence of an acidic medium. This method also produces aniline as a side-product. The advantage of a reduction using a noble metal catalyst is that it involves a single step reaction, an environment friendly and more efficient process, as there is no evolution of an environmentally harmful gas. Moreover, the side-product aniline is also a valuable chemical.

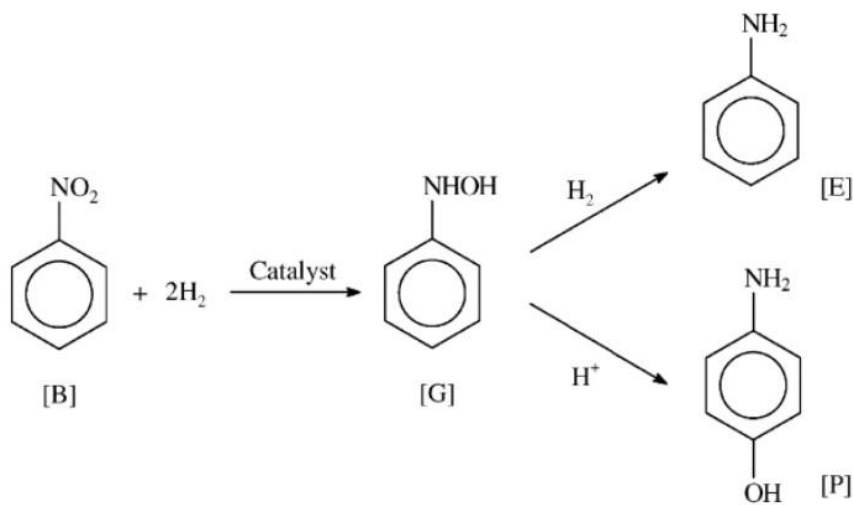

Nitrobenzene is reduced to the intermediate β-phenylhydroxylamine (PHA) followed by the in situ rearrangement to PAP in presence of acidic medium. This is a multiphase catalytic reaction. The phases involved are solid (catalyst), gaseous ($H_2$), organic phase (nitrobenzene) and the aqueous acidic phase. The overall rate of reaction is dependent on the gas-liquid, liquid-liquid and liquid-solid mass transfer and interfacial areas. The kinetics and rate data for this reaction has been investigated using a batch slurry reactor (C.V. Rode). The reaction to the intermediate β-phenylhydroxylamine is instantaneous, since β-phenylhydroxylamine is not detected in the product. The overall reaction scheme can be represented as follows

$$A + 3B \rightarrow E + 2W$$

$$A + 2B \rightarrow P + W$$

where A = Nitrobenzene, B = Hydrogen, E = Aniline, P = p-aminophenol, W = Water

On adding both the reactions we get

$$2A + 5B \rightarrow E + P + 3W$$

The effect of hydrogen pressure over nitrobenzene conversion and the selectivity has been studied (Yingxin, Liu) extensively. The optimum combination of selectivity and conversion is obtained as 61% conversion of nitrobenzene and PAP selectivity of 77.8%. The hydrogen pressure was 0.4 MPa and the reaction temperature of $85°C$ was determined for best possible products. We decided to use these results in flowsheet simulation. A stoichiometric reactor having nitrobenzene conversion of 60% and selectivity of 70 % was modelled to produce the desired products.

## 2.3 CATALYST PROPERTIES

The catalyst used for this process is the Pt-C in acidic medium. Generally the noble metal i.e. Pt is used with a supported catalyst. The content of platinum in the supported catalyst ranges between 0.01 to 10% by weight of the total weight of the supported catalyst.[4] The catalyst used for preparing p-aminophenol from nitrobenzene uses a mixture of zirconium sulphate with a supported hydrogenation noble metal.[4] Here, there are two competing reactions, one is the hydrogenation reaction producing *p*-aminophenol and the other is the isomerization reaction producing the side-product aniline. If the amount of Pt taken is very high, then the reaction will hydrogenate the intermediate phenylhydroxylamine (PHA) to aniline. If the hydrogenation of nitrobenzene (NB) to *p*-aminophenol (PAP) is carried out using Pt as a hydrogenation noble metal, the commonly used catalyst are $Zr(SO_4)_2(H_2O)_4$ + Pt/C, $Zr(SO_4)_2(H_2O)_4$ + Pt/ZrS, $Zr(SO_4)_2(H_2O)_4$ + PtVZrO$_2$, $Zr(SO_4)_2(H_2O)_4$ + PVTiO$_2$, $Zr(SO_4MH_2O)_4$ + Pt/Al$_2$O$_3$, $Zr(SO_4)_2(H_2O)_4$ + PVSiO$_2$, $Zr(SO_4)_2(H_2O)_4$ + Pt/MgLaO.[4] The weight ratio of Zirconium Sulphate to hydrogenation noble metal may vary in great proportion. After thorough literature study, we have selected the weight ratio of Zirconium Sulphate to Pt as 50:1. The weight of Pt is kept as low as possible, so that the side reaction producing aniline can be reduced.

## 2.4 HYDROGEN GENERATION

We have taken Hydrogen directly for this process, but it can also be manufactured in the same plant through the Hydrocarbon Steam-Reforming method, which is used to produce most of the industrial Hydrogen.

Natural Gas contains methane which can be used to produce hydrogen via steam-methane reforming. Natural gas contains more than 90% of methane, which will be a major source of Hydrogen. High-pressure steam is reacted with natural gas in the presence of catalysts producing hydrogen, carbon monoxide and a trace amount of carbon-dioxide. This method also involves the water-gas shift reaction between CO and H$_2$O.

Steam-methane reforming

$$CH_4 + H_2 \rightarrow CO + 3H_2O$$

Water-gas shift reaction

$$CO + H_2O \rightarrow CO_2 + H_2$$

A cleaner way to produce hydrogen is through the electrolysis of water. Through electrolysis Hydrogen is produced at cathode and Oxygen is produced at anode.

$$2H^+ + 2e^{-1} \rightarrow H_2 \qquad \text{Cathode}$$

$$2OH^{-1} \rightarrow H_2O + 1/2O_2 + 2e^{-1} \qquad \text{Anode}$$

Hydrocarbon Steam-Reforming is the cheapest way of producing industrial hydrogen. During the initial years of operation, hydrogen can be produced using Hydrocarbon-Steam reforming. With the passage of time electrolysis can be used instead of steam reforming.

## 2.5 SEPARATION OF PARA-AMINOPHENOL

The product stream form the reactors contains the unreacted nitrobenzen, hydrogen, water, aniline and *p*-aminophenol. Since hydrogen is a highly volatile gas it can be easily separated using a flash. In the next stage a distillation column is used to separate out the *p*-aminophenol (PAP). PAP is least volatile compound in the column (BP 284°C), so it is collected at the bottom of the first distillation column. The stream at the top of the column is again passed through a distillation column to separate out unreached nitrobenzene and the by-product aniline. Aniline mixed with water is obtained from the second column, which is further purified using a decanter, and thus the products and by-products are separated out.

# 3. ASSEMBLY OF DATABASE

## 3.1 COST OF CHEMICALS

Raw materials used in the process are Nitrobenzene and Hydrogen. Cost of Nitrobenzene is Rs 70/kg and the cost of the Hydrogen is Rs 95/kg. Feed condition is 30 degree celsius at 1 atm. Products by the reaction of the raw material in the reactor are p-aminophenol, aniline, water with some amount of unreacted hydrogen and nitrobenzene.

## 3.2 COST OF CATALYST

The catalyst used for the conversion of nitrobenzene to *p*-aminophenol is the Pt-C supported over hydrated zirconium sulphate i.e Pt/C with sulphuric acid providing the acidic medium. The weight ratio of the catalyst matrix and platinum is 50:1. Therefore the percentage of Pt loading in the catalyst is approximately 2%. The sulphuric acid concentration was 98% purity. The selectivity of producing *p*-aminophenol decreases with increasing Pt concentration. Weight ratio of 50:1 will provide a high selectivity.

| CATALYST COST | | | |
|---|---|---|---|
| Catalyst | Price (per kg) | Quantity (kg per annum) | Cost (per annum) |
| Pt/C catalyst | Rs 1000 | 2400 | 2400000 |
| Sulphuric acid (98%) | Rs 4.8 | 120000 | 576000 |

## 3.3 CHEMICAL PROPERTIES

The table describes the basic chemical properties of the chemicals present or produced during the process.[10][11]

| CHEMICAL PROPERTIES | | | | |
|---|---|---|---|---|
| Chemical Name | Molar Mass (gm/mol) | Density (gm/cm³) | Boiling Point(°C) | Heat Capacity (cal/mol) |
| Nitrobenzene | 123.11 | 1.2 | 210.9 | 44 |
| Hydrogen | 2.016 | 0.089 (gm/L) | -252.9 | 6.85 |
| Water | 18.01 | 0.998 | 100 | 1 |
| Aniline | 93.13 | 1.02 | 184.1 | 46.2 |
| Para-Aminophenol | 109.13 | 1.13 | 284 | 45.6 |

## 3.4 PLANT CAPACITY CALCULATIONS

We have assumed the capacity of the plant to be 18,000 tonnes per annum (TPA) and considered that the plant functions for 24 hours for 300 working days in an year. Therefore, we can calculate the hourly capacity of the plant as follows

Para-Amino Phenol Production = 18,000 Tonnes per Annum

$$= \frac{18,000 \times 1000 \ kg}{300 \ days}$$
$$= \frac{18,000 \times 1000 \ kg}{300 \times 24 \ hours}$$
$$= 2500 \ kg/hr$$

Converting into moles, we get the production rate of p-aminophenol as 22.94 kmol/hr. In actual simulation, we get 22.7 kmol/hr of p-aminophenol.

## 4. PFD AND MATERIAL BALANCES

### 4.1 PROCESS FLOW DIAGRAM

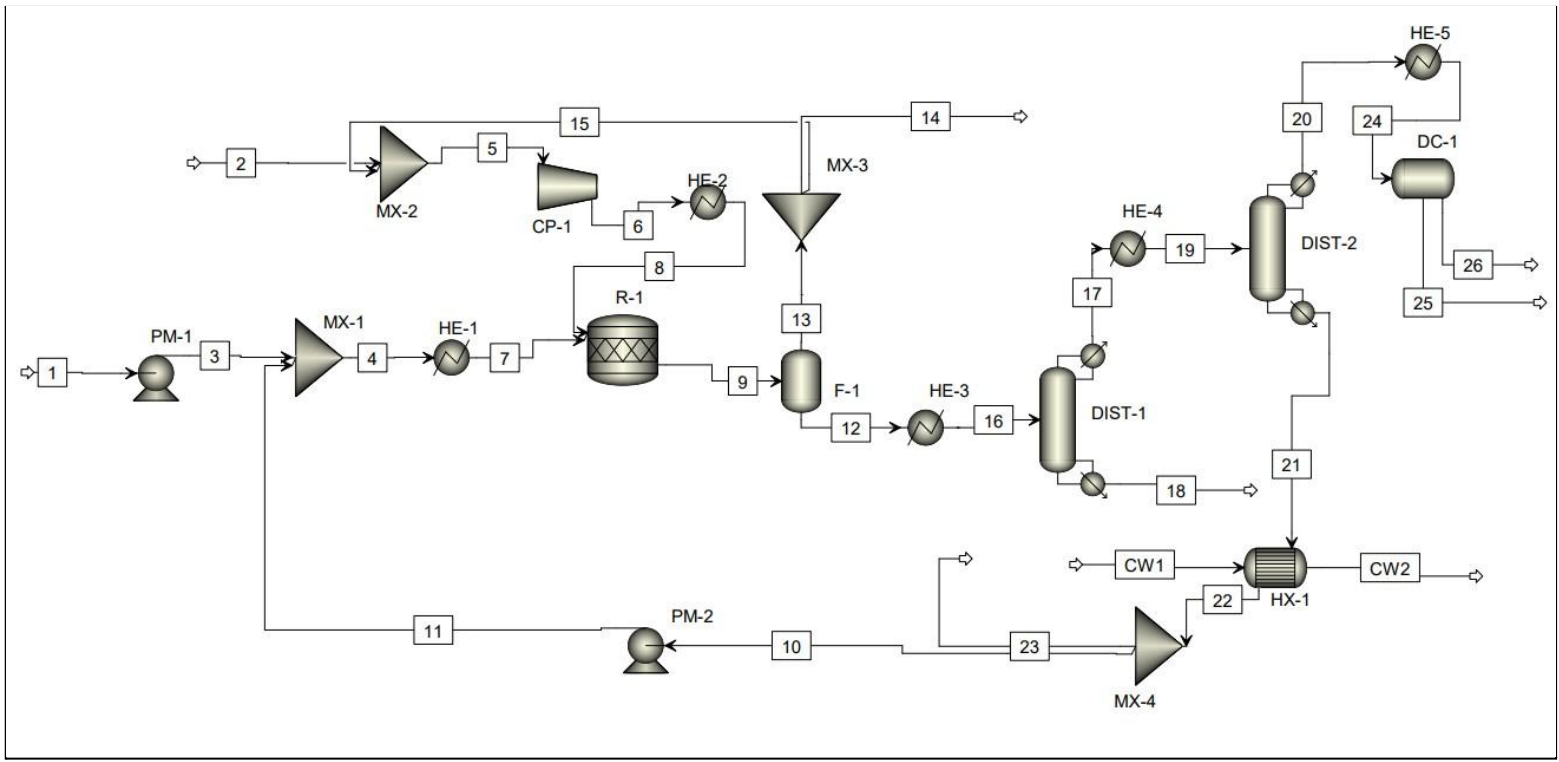

### 4.2 OVERALL MATERIAL BALANCES

→STREAM 1 to 8

|  | 1 | 2 | 3 | 4 | 5 | 6 | 7 | 8 |
|---|---|---|---|---|---|---|---|---|
| Phase | Liquid | Vapor | Liquid | Liquid | Vapor | Vapor | Liquid | Vapor |
| Temperature(°C) | 25.000 | 30.00 | 25.100 | 47.791 | 27.725 | 246.765 | 85.000 | 85.00 |
| Pressure (bar) | 1.000 | 1.000 | 4.000 | 4.000 | 1.000 | 4.000 | 4.000 | 4.000 |
| Mass Flows (kg/hr) | 4308.893 | 187.48 | 4308.9 | 6813.8 | 400.431 | 400.431 | 6813.84 | 400.4 |
| Mole Flows (kmol/hr | 35.000 | 93.00 | 35.000 | 55.738 | 169.933 | 169.933 | 55.738 | 169.9 |
| Component wise Material Balance for each stream (kmol/hr) | | | | | | | | |
| Nitrobenzene | 35.000 | 0.000 | 35.000 | 54.082 | 0.006 | 0.006 | 54.082 | 0.006 |
| Hydrogen | 0.000 | 93.000 | 0.000 | 0.000 | 166.411 | 166.411 | 0.000 | 166.4 |
| P-Aminophenol | 0.000 | 0.000 | 0.000 | 0.097 | 0.000 | 0.000 | 0.097 | 0.000 |
| Water | 0.000 | 0.000 | 0.000 | 0.000 | 3.504 | 3.504 | 0.000 | 3.504 |
| Aniline | 0.000 | 0.000 | 0.000 | 1.558 | 0.011 | 0.011 | 1.558 | 0.011 |

→STREAM 9 to 15

| | 9 | 10 | 11 | 12 | 13 | 14 | 15 |
|---|---|---|---|---|---|---|---|
| Phase | Mixed | Liquid | Liquid | Liquid | Vapor | Vapor | Vapor |
| Temperature (C) | 85.000 | 85.000 | 85.115 | 25.000 | 25.00 | 25.00 | 25.00 |
| Pressure (bar) | 4.000 | 1.000 | 4.000 | 1.000 | 1.000 | 1.000 | 1.000 |
| Mass Flows (kg/hr) | 7214.285 | 2504.95 | 2504.95 | 6948.1 | 266.2 | 53.24 | 212.95 |
| Mole Flows (kmol/hr) | 193.217 | 20.738 | 20.738 | 97.051 | 96.17 | 19.23 | 76.933 |
| **Component wise Material Balance for each stream (kmol/hr)** | | | | | | | |
| Nitrobenzene | 21.635 | 19.082 | 19.082 | 21.627 | 0.008 | 0.002 | 0.006 |
| Hydrogen | 91.769 | 0.000 | 0.000 | 0.005 | 91.76 | 18.35 | 73.411 |
| p-aminophenol | 22.815 | 0.097 | 0.097 | 22.815 | 0.000 | 0.000 | 0.000 |
| water | 45.693 | 0.000 | 0.000 | 41.313 | 4.380 | 0.876 | 3.504 |
| aniline | 11.305 | 1.558 | 1.558 | 11.291 | 0.014 | 0.003 | 0.011 |

STREAM 16 to 22

| | 16 | 17 | 18 | 19 | 20 | 21 | 22 |
|---|---|---|---|---|---|---|---|
| Phase | Mixed | Vapor | Liquid | Mixed | Vapor | Liquid | Liquid |
| .Temperature (°C) | 120.00 | 176.91 | 283.48 | 120.00 | 134.48 | 208.95 | 85.000 |
| Pressure (bar) | 1.000 | 1.000 | 1.000 | 1.000 | 1.000 | 1.000 | 1.000 |
| Mass Flows (kg/hr) | 6948.1 | 4468.2 | 2479.9 | 4468.2 | 1684.9 | 2783.3 | 2783.3 |
| Mole Flows (kmol/hr) | 97.051 | 74.328 | 22.723 | 74.328 | 51.286 | 23.042 | 23.042 |
| **Component wise Material Balance for each stream (kmol/hr)** | | | | | | | |
| Nitrobenzene | 21.627 | 21.612 | 0.015 | 21.612 | 0.410 | 21.203 | 21.203 |
| Hydrogen | 0.005 | 0.005 | 0.000 | 0.005 | 0.005 | 0.000 | 0.000 |
| p-aminophenol | 22.815 | 0.108 | 22.707 | 0.108 | 0.000 | 0.108 | 0.108 |
| water | 41.313 | 41.313 | 0.000 | 41.313 | 41.313 | 0.000 | 0.000 |
| aniline | 11.291 | 11.290 | 0.000 | 11.290 | 9.559 | 1.731 | 1.731 |

STREAM 23 onwards

| | 23 | 24 | 25 | 26 | CW1 | CW2 |
|---|---|---|---|---|---|---|
| Phase | Liquid | Mixed | Liquid | Liquid | Liquid | Mixed |
| Temperature (C) | 85.000 | 30.000 | 30.000 | 30.000 | 5.000 | 102.083 |

| Pressure (bar) | 1.000 | 1.000 | 1.000 | 1.000 | 1.000 | 1.000 |
|---|---|---|---|---|---|---|
| Mass Flows (kg/hr) | 278.327 | 1684.9 | 967.55 | 717.36 | 360.30 | 360.306 |
| Mole Flows (kmol/hr) | 2.304 | 51.286 | 11.470 | 39.817 | 20.000 | 20.000 |
| **Component wise Material Balance for each stream (kmol/hr)** | | | | | | |
| Nitrobenzene | 2.120 | 0.410 | 0.410 | 0.000 | - | - |
| Hydrogen | 0.000 | 0.005 | 0.005 | 0.000 | - | - |
| p-aminophenol | 0.011 | 0.000 | 0.000 | 0.000 | - | - |
| water | 0.000 | 41.313 | 1.497 | 39.816 | 20.000 | 20.000 |
| aniline | 0.173 | 9.559 | 9.558 | 0.001 | - | - |

## 5. UTILITY REQUIREMENTS

A summary of the amount of electricity, cooling water, and steam required to operate the plant can be found in the table below.

| UTILITY | | | |
|---|---|---|---|
| Name | Fluid | Rate | Rate Units |
| Electricity | | 374.203 | KW |
| Cooling Water | Water | 0.085646 | MMGAL/H |
| Steam @100PSI | Steam | 3.011514 | KLB/H |
| Steam @400PSI | Steam | 5.362324 | KLB/H |

# 6. PROCESS DESCRIPTION AND EQUIPMENT DETAILS

## 6.1 SIMPLIFIED BLOCK DIAGRAM

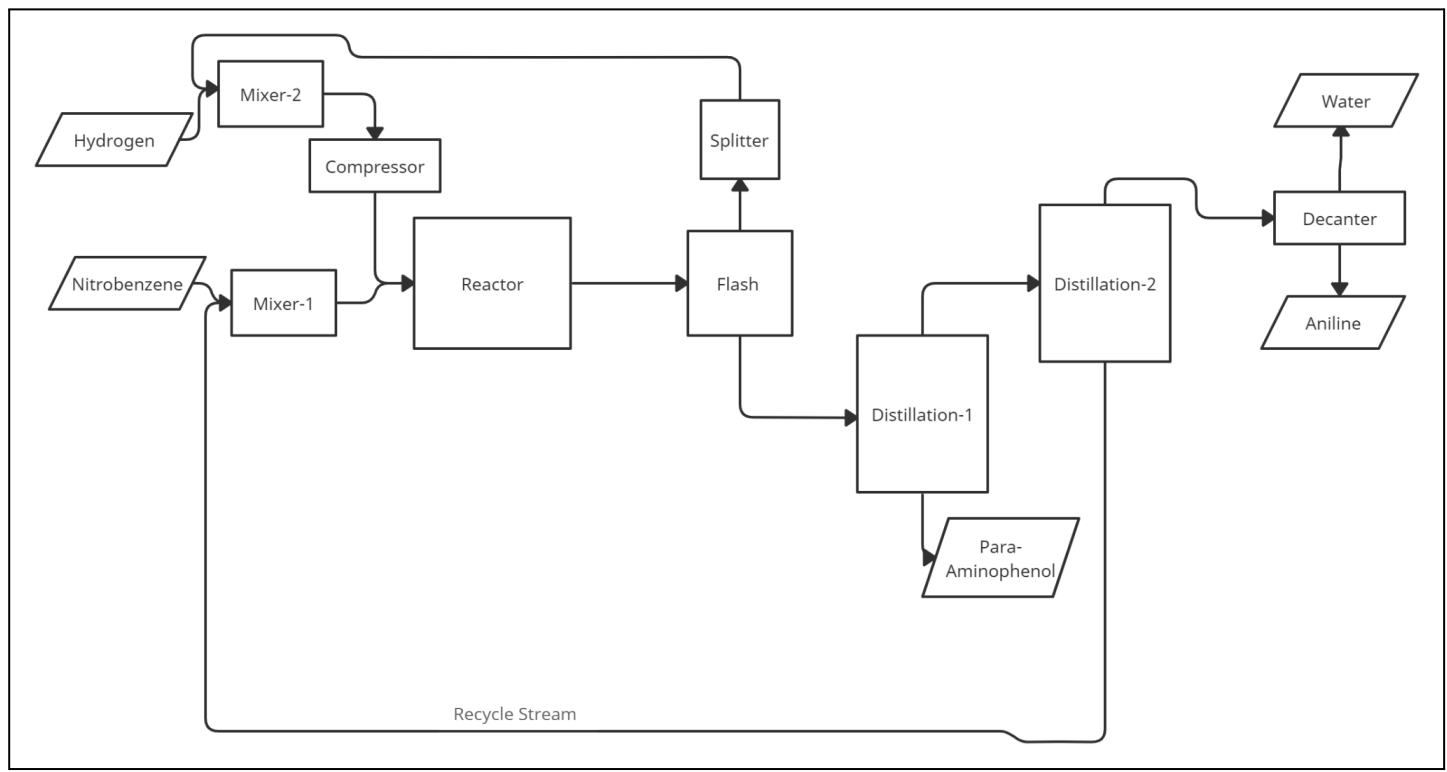

According to the route selected for the production of the p-aminophenol, raw materials required are Nitrobenzene and Hydrogen. Input stream is fed into the stoichiometric reactor with overall conversion of nitrobenzene as 0.60 and the selectivity of p-aminophenol as 0.70. The output stream of the reactor consists of nitrobenzene, hydrogen, aniline, water and para-aminophenol. Based on the compound's chemical properties, (as mentioned in section 4.3) first we have used a flash drum to separate hydrogen from the remaining compounds of the output stream. Hydrogen comes from the top stream of the flash drum. Hydrogen stream is recycled, with 0.8 split fraction for recycle stream. Input stream and recycled hydrogen stream are mixed and compressed before entering the reactor. The bottom stream of the flash drum consists of nitrobenzene, aniline, water and para aminophenol. This stream is fed to the first distillation column, operating at 1 bar partial condenser with a reflux ratio of 1.5. In this distillation column, p-aminophenol is separated from the mixture as the bottom product. The distillate is fed into another distillation column, operating at 1 bar partial condenser with a reflux ratio 1.25 where nitrobenzene is separated from the bottom stream. This nitrobenzene is recycled, the input and recycle stream are mixed in the mixer and then fed into the reactor. The top stream of the second distillation column consists of water and aniline. This stream is fed into the decanter, where water and aniline are separated. Hence with the use of a flash drum, two-distillation column and a decanter all the products from the reactor are separated.

## 6.2 EQUIPMENTS USED

**a.     Reactor**
Stoichiometric reactor has been used in the ASPEN Simulation where the nitrobenzene and hydrogen are reacted to obtain the desired product i.e. para-aminophenol.

| Reactor | |
|---|---|
| Liquid volume [l] | 889.605 |
| Vessel diameter [meter] | 0.610 |
| Vessel tangent to tangent height [meter] | 3.048 |
| Design gauge pressure [barg] | 4.710 |
| Design temperature [C] | 121.111 |

**b.     Flash Drum**

Flash Drum is used to separate the hydrogen from the output stream of the reactor. Output stream consists of water, aniline, para- aminophenol, and unreacted hydrogen and nitrobenzene.

| F-1-flash drum | |
|---|---|
| Liquid volume [l] | 3269.297609 |
| Vessel diameter [meter] | 1.0668 |
| Vessel tangent to tangent height [meter] | 3.6576 |
| Design gauge pressure [barg] | 4.710460804 |
| Design temperature [C] | 121.1111111 |
| Operating temperature [C] | 85 |

**c.     Distillation Column**

Two distillation columns are used to separate the para-aminophenol and nitrobenzene respectively.

| | DIST-1-tower | DIST-2-tower |
|---|---|---|
| Diameter Bottom section [meter] | 1.524 | 1.372 |
| Bottom tangent to tangent height [meter] | 8.534 | 10.973 |
| Design gauge pressure Bottom [barg] | 1.034 | 1.034 |
| Design temperature Bottom [C] | 311.264 | 236.729 |
| Operating temperature Bottom [C] | 283.487 | 208.952 |
| Number of trays Bottom section | 8.000 | 12.000 |
| Bottom Tray type | SIEVE | SIEVE |
| Bottom Tray spacing [meter] | 0.610 | 0.610 |
| Reflux Ratio | 1.5 | 1.25 |

**d.     Decanter**

Decanter is used in the final stage to separate water and aniline.

| | DC-1 |
|---|---|
| Liquid volume [l] | 2401.932948 |
| Vessel diameter [meter] | 0.9144 |
| Vessel tangent to tangent height [meter] | 3.6576 |
| Design gauge pressure [barg] | 1.03425 |
| Vacuum design gauge pressure [barg] | -1.00667 |
| Design temperature [C] | 121.1111111 |
| Operating temperature [C] | 30 |

e. **Compressor**

Compressor is used to compress the output stream of the mixer, which consists of the input hydrogen and the recycled hydrogen.

| Centrif gas compressor | |
|---|---|
| | **CP-1** |
| Actual gas flow rate Inlet [l/min] | 70884.03 |
| Design gauge pressure Inlet [barg] | -0.01 |
| Design temperature Inlet [C] | 27.73 |
| Design temperature Outlet [C] | 246.77 |
| Design gauge pressure Outlet [barg] | 2.99 |
| Driver power [kW] | 303.04 |
| Molecular weight | 2.36 |
| Specific heat ratio | 1.40 |
| Compressibility factor Inlet | 1.00 |
| Compressibility factor Outlet | 1.00 |
| Driver type | MOTOR |

f. **Heaters**

Five heaters are used in the simulation to control the temperature and pressure of the streams entering into the distillation columns, reactor or the flash drum.

| | Heater | | | | |
|---|---|---|---|---|---|
| | **HE-1** | **HE-2** | **HE-3** | **HE-4** | **HE-5** |
| Quantity | 1.00 | 1.00 | 1.00 | 1.00 | 1.00 |
| Heat transfer area [sqm] | 1.50 | 7.52 | 9.25 | 7.00 | 48.23 |
| Tube design gauge pressure [barg] | 7.61 | 4.16 | 7.61 | 4.16 | 4.16 |
| Tube design temperature [C] | 192.11 | 274.54 | 192.11 | 204.69 | 162.26 |
| Tube operating temperature [C] | 164.33 | 35.00 | 164.33 | 35.00 | 35.00 |
| Tube outside diameter [meter] | 0.03 | 0.03 | 0.03 | 0.03 | 0.03 |
| Shell design gauge pressure [barg] | 4.73 | 4.71 | 4.73 | 2.43 | 2.43 |
| Shell design temperature [C] | 121.11 | 274.54 | 147.78 | 204.69 | 162.26 |
| Shell operating temperature [C] | 85.00 | 246.77 | 120.00 | 176.91 | 134.48 |
| Tube length extended [meter] | 6.10 | 6.10 | 6.10 | 6.10 | 6.10 |
| Tube pitch [meter] | 0.03 | 0.03 | 0.03 | 0.03 | 0.03 |
| Number of tube passes | 1.00 | 1.00 | 1.00 | 1.00 | 1.00 |
| Number of shell passes | 1.00 | 1.00 | 1.00 | 1.00 | 1.00 |
| Heat Duty[cal/s] | 27760.39 | -53583.8 | 159599.2 | -143220 | -165063 |

### g. Heat Exchangers

The bottom steam coming from the second column continues nitrobenzene. This stream needs to be cooled before sending to the reactor using a heat exchanger.

| | HX-1 |
|---|---|
| Heat transfer area [sqm] | 4.435531 |
| Tube design gauge pressure [barg] | 0.020961 |
| Tube design temperature [C] | 236.7294 |
| Tube operating temperature [C] | 102.0834 |
| Tube outside diameter [meter] | 0.0254 |
| Shell design gauge pressure [barg] | 0.020961 |
| Shell design temperature [C] | 236.7294 |
| Shell operating temperature [C] | 208.9516 |
| Tube length extended [meter] | 6.096 |
| Tube pitch [meter] | 0.03175 |

## 6.3 SELECTIVITY AND CONVERSION IN RSTOIC

The optimum conversion achievable using Pt-C catalyst in acidic medium is 61% with a selectivity of 77.8%[2]. Since, these optimum values were obtained in laboratory settings in batch reactors, the actual conversion and selectivity would be lesser.

We had set the conversion and selectivity to be 60% and 70% respectively. The actual conversion and selectivity in the plant reactor will be approximately similar, as the reactor conditions are the same. To produce *p-aminophenol*, we used a stoichiometric reactor having two reactions for the desired and the undesired products.

The overall reaction scheme can be represented as -

$$A + 2B \rightarrow P + W \quad (desired)$$

$$A + 3B \rightarrow E + 2W \quad (undesired)$$

where A = Nitrobenzene, B = Hydrogen, E = Aniline, P = *p*-aminophenol, W = Water

Let $\alpha$ and $\beta$ be the extent of the desired and the undesired reaction respectively. Therefore the composition both the reactions can be expressed using the concept of extent of reaction as

$$A + 2B \rightarrow P + W \quad (desired)$$

At completion of reaction, $(F_{Ao} - \alpha) + (2F_{Bo} - 2\alpha) \rightarrow \alpha + \alpha$

$$A + 3B \rightarrow E + 2W \quad (undesired)$$

$$(F_{Ao} - \beta) + (3B - 3\beta) \rightarrow \beta + 2\beta$$

The optimum selectivity of 75.5 % is required. Therefore the proportion of desired product is

$$\therefore \frac{\alpha}{\alpha+\beta} \times 100 = 70$$

Also, the overall conversion of 61% is selected.

$$\therefore Conversion(X) = \frac{moles\ of\ A\ reacted}{moles\ of\ A\ feed} = \frac{\alpha+\beta}{2F_{Ao}} = 0.60$$

Solving for α and β we get

$\alpha = 0.84\ F_{Ao}$ and $\beta = 0.36\ F_{Ao}$

The conversions for individual reactions can be calculated as

$$X_1 = \frac{\alpha}{2F_{Ao}} = \frac{0.84\ F_{Ao}}{2F_{Ao}} = 0.42 \qquad X_2 = \frac{\beta}{2F_{Ao}} = \frac{0.36\ F_{Ao}}{2F_{Ao}} = 0.18$$

A stoichiometric reactor operating at $85°C$ and 0.4 MPa having the conversion of 0.42 and 0.18 for the desired and undesired reactions was modelled to produce the product.

The overall reaction scheme can be represented as -

$$A + 2B \rightarrow P + W \qquad (desired)$$

$$A + 3B \rightarrow E + 2W \qquad (undesired)$$

where A = Nitrobenzene, B = Hydrogen, E = Aniline, P = p-aminophenol, W = Water

Let α and β be the extent of the desired and the undesired reaction respectively. Therefore the composition both the reactions can be expressed using the concept of extent of reaction as

$$A + 2B \rightarrow P + W \qquad (desired)$$

At completion of reaction, $(F_{Ao} - \alpha) + (2F_{Bo} - 2\alpha) \rightarrow \alpha + \alpha$

$$A + 3B \rightarrow E + 2W \qquad (undesired)$$

$$(F_{Ao} - \beta) + (3B - 3\beta) \rightarrow \beta + 2\beta$$

The optimum selectivity is 75.5 % is required. Therefore the proportion of desired product is

$$\therefore \frac{\alpha}{\alpha+\beta} \times 100 = 70$$

Also, the overall conversion of 61% is selected.

$$\therefore Conversion(X) = \frac{moles\ of\ A\ reacted}{moles\ of\ A\ feed} = \frac{\alpha+\beta}{2F_{Ao}} = 0.60$$

Solving for α and β we get

$\alpha = 0.84\ F_{Ao}$ and $\beta = 0.36\ F_{Ao}$

The conversions for individual reactions can be calculated as

$$X_1 = \frac{\alpha}{2F_{Ao}} = \frac{0.84\ F_{Ao}}{2F_{Ao}} = 0.42 \qquad X_2 = \frac{\beta}{2F_{Ao}} = \frac{0.36\ F_{Ao}}{2F_{Ao}} = 0.18$$

A stoichiometric reactor operating at $85°C$ and 0.4 MPa having the conversion of 0.42 and 0.18 for the desired and undesired reactions was modelled to produce the product.

## 6.4 PRESSURE VESSEL DESIGN

A high pressure vessel should be constructed of appropriate material and should be of sufficient thickness so that it is able handle the pressure inside it. Moreover, while fabricating the vessel the maximum limit of the joints and welds must be examined thoroughly.

The thickness of a pressure vessel is given as

$$t = \frac{PD_i}{2f_{ds}J - P}$$

where $t$ is the shell thickness, $P$ is design pressure, $f_{ds}$ is design stress, $D_i$ is internal diameter and $J$ is the joint efficiency.

It is safe to consider the design pressure to be 10% more than the maximum pressure.

From the simulation properties of reactor are

$D_i = 0.610$ m     $P_{des} = 5.710$ bar

Assuming the design stress $f_{ds}$ to be 5,000 psi (344.7 bar) (from literature)

$$t = \frac{5.710 \times 0.610}{2 \times 344.7 \times 1 - 5.710} = 5.09 \times 10^{-3}\ m \qquad \therefore t = 5.09\ mm$$

**Stress due to internal pressure**

Circumferential stress (tensile), $f_C = \frac{PD_i}{2t} = \frac{5.71 \times 10^4 \times 0.610}{2 \times 5.09 \times 10^{-3}} = 34.22\ MN/m^2$

Axial stress (compressive) $f_a = \frac{PD_i}{4t} = \frac{f_C}{2} = 17.11\ MN/m^2$

**Stress due to weight** (compressive), $f_w = \frac{W}{\pi(D_i + t)t}$

An approximate expression for the total weight of the shell is

$$W_{shell} = C_v \pi \rho_m D_m g (H_V + 0.8 D_m) t \times 10^{-3}$$

where  $H_v$ = of shell or reactor = 3.048 m

$\rho_m$ = density of vessel material = 7800 $kg/m^3$ (ASTM A516 Grade 70 steel)

$t$ = wall thickness = 5.09 mm

$g$ = gravitational acceleration, 9.81 $m/s^2$

$D_m = \left(D_i + t \times 10^{-3}\right) = 0.61509$ m

$C_v$ = factor taken to account the weight of manhole, internal support etc., taking

it to be 1.08

$$\therefore W_{shell} = 1.08\pi \times 7800 \times 0.615 \times 9.81 \times (3.54)5.09 \times 10^{-3} = 2877 \, N$$

$$\therefore f_w = \frac{2877}{\pi (0.61) 0.00509} = 0.295 \, MN/m^2$$

The diameter of our reactor vessel is 0.62 m, for a satisfactory design the thickness of the reactor vessel should be larger than 3 mm. We have selected the thickness to be 5.09 mm, which is more than the threshold.

## 7. EQUIPMENT COST SUMMARY

→**Reactor**

| | | | | | | |
|---|---|---|---|---|---|---|
| | | | Reactor | | | |
| Name | Equipment Cost [USD] | Installed Cost [USD] | Equipment Weight [LBS] | Installed Weight | Utility Cost [USD/HR] | Total Utility Cost(USD/Year) |
| R-1 | 64500 | 205100 | 5500 | 20506 | 0 | 0 |

J→**Flash Drum**

| | | | | | | |
|---|---|---|---|---|---|---|
| | | | Flash Drum | | | |
| Name | Equipment Cost [USD] | Installed Cost [USD] | Equipment Weight [LBS] | Installed Weight | Utility Cost [USD/HR] | Total Utility Cost(USD/Year) |
| F-1 | 17700 | 106700 | 3000 | 13409 | 0 | 0 |

→**Distillation Column**

| | | | | | | |
|---|---|---|---|---|---|---|
| | | | Distillation Column | | | |
| Name | Equipment Cost [USD] | Installed Cost [USD] | Equipment Weight [LBS] | Installed Weight | Utility Cost [USD/HR] | Total Utility Cost(USD/Year) |
| DIST-1 | 206700 | 678800 | 54990 | 126898 | 4.554475 | 32792.22 |
| DIST-2 | 154100 | 563300 | 31070 | 85153 | 65.258894 | 469864.0368 |

→**Decanter**

| | | | | | | |
|---|---|---|---|---|---|---|
| | | | Decanter | | | |
| Name | Equipment Cost [USD] | Installed Cost [USD] | Equipment Weight [LBS] | Installed Weight | Utility Cost [USD/HR] | Total Utility Cost(USD/Year) |
| DC-1 | 15800 | 110500 | 2600 | 11524 | 0 | 0 |

→**Compressor**

| | | | | | | |
|---|---|---|---|---|---|---|
| | | | Compressor | | | |
| Name | Equipment Cost [USD] | Installed Cost [USD] | Equipment Weight [LBS] | Installed Weight | Utility Cost [USD/HR] | Total Utility Cost(USD/Year) |
| CP-1 | 1374000 | 1521800 | 22600 | 38729 | 26.01675 | 187320.6 |

→**Heater**

| Heater | | | | | | |
|---|---|---|---|---|---|---|
| Name | Equipment Cost [USD] | Installed Cost [USD] | Equipment Weight [LBS] | Installed Weight | Utility Cost [USD/HR] | Total Utility Cost(USD/Year) |
| HE-1 | 8400 | 62300 | 500 | 7792 | 3.632125 | 26151.3 |
| HE-2 | 10100 | 62400 | 1200 | 7741 | 0.5508 | 3965.76 |
| HE-3 | 10800 | 65700 | 1300 | 8973 | 20.881599 | 150347.5128 |
| HE-4 | 10000 | 61400 | 1100 | 7291 | 1.47216 | 10599.552 |
| HE-5 | 17300 | 88200 | 4100 | 16651 | 1.69668 | 12216.096 |

→**Mixer**

| Mixer | | | | | | |
|---|---|---|---|---|---|---|
| Name | Equipment Cost [USD] | Installed Cost [USD] | Equipment Weight [LBS] | Installed Weight | Utility Cost [USD/HR] | Total Utility Cost(USD/Year) |
| MX-1 | 666.67 | 120 | 0 | 0 | 1 | 7200 |
| MX-2 | 666.67 | 120 | 0 | 0 | 1 | 7200 |
| MX-3 | 666.67 | 120 | 0 | 0 | 1 | 7200 |

→**Pump**

| Pump | | | | | | |
|---|---|---|---|---|---|---|
| Name | Equipment Cost [USD] | Installed Cost [USD] | Equipment Weight [LBS] | Installed Weight | Utility Cost [USD/HR] | Total Utility Cost(USD/Year) |
| PM-1 | 4000 | 31800 | 200 | 3130 | 0.0434 | 312.48 |
| PM-2 | 4000 | 31800 | 190 | 3120 | 0.028675 | 206.46 |

→**Heat Exchanger**

| Heat Exchanger | | | | | | |
|---|---|---|---|---|---|---|
| Name | Equipment Cost [USD] | Installed Cost [USD] | Equipment Weight [LBS] | Installed Weight | Utility Cost [USD/HR] | Total Utility Cost(USD/Year) |
| HX-1 | 9700 | 64400 | 1000 | 8517 | 0 | 0 |

**NET COST**

| Net Cost | | | | | | |
|---|---|---|---|---|---|---|
| (In Rs) | Equipment Cost | Installed Cost | Equipment Weight | Installed Weight | Utility Cost(Rs/HR) | Total Utility Cost(per Year) |
| Total Cost | Rs14.38 crore | Rs 27.56 crore | 58207.5 Kg | 161745.3 Kg | 9363.54 | Rs 6.74 crore |

The cost of the Equipment and its installation is Rs 41.94 crore

## 8. FIXED COST SUMMARY

In order to consider the total permanent investestment, both direct and indirect components must be considered. Direct cost includes the cost of land, equipment cost, installation, piping, electrical system and building infrastructure around the plan are considered. Indirect components include the salary of the engineers and supervisors, appointed for the functioning and maintaining the plant.

### 8.1 Direct Cost

| Direct Cost | |
|---|---|
|  | INR (crores) |
| Purchased Equipment | 14.38 |
| Installation | 27.56 |
| Instrumentation and control | 22.11 |
| Piping | 39.87 |
| Electrical system | 7.45 |
| Building | 11.55 |
| Yard Improvement | 6.86 |
| Service Facility | 39.11 |
| Land | 12.00 |
| Packaging and Storage | 8 |
| Total Cost | **Rs 188.91 crore** |

### 8.2 Indirect Cost

| MANPOWER REQUIREMENTS (PERMANENT) | | | |
|---|---|---|---|
|  | Number | Salary (INR) | Total yearly Amount |
| Engineer | 12 | 8 LPA | 9600000 |
| Administration | 12 | 6 LPA | 7200000 |
| Logistics & Inventory Management | 10 | 6 LPA | 6000000 |
| Manager / Supervisor | 6 | 12 LPA | 7200000 |
| TOTAL | 40 |  | **Rs 3,00,00,000 crore** |

The net fixed cost is the sum of the direct and indirect fixed cost. Hence, the total fixed cost for our plant is **Rs. 191.91 crores.**

## 9. OPERATING COST

Operating cost is the cost required to run the plant, which includes the cost of raw material, labours, electricity, water, maintenance and insurance. The estimate below is on a yearly basis.

a. **Raw Material and Catalyst**

| Raw Material and Catalyst | | | | |
|---|---|---|---|---|
| | **Cost per Kg** | Rate | Quantity(kg per annum) | Cost per Annum |
| Nitrobenzene | 70 | 35 kmol/hr | 30996000 | 2169720000 |
| Hydrogen | 95 | 93 kmol/hr | 1339200 | 127224000 |
| Pt/C catalyst | 1000 Rs/kg | | 2400 | 2400000 |
| Sulphuric acid | 98% Rs 4.8 kg | | 120000 | 576000 |
| | | | **Total Cost** | **Rs. 229.99 crore** |

b.  **Utility**

| Utility | | | | | |
|---|---|---|---|---|---|
| Name | Fluid | Rate | Rate Units | Cost per Hour(USD /H | Year Cost in Rs |
| Electricity | | 374.203 | KW | 29.000732 | 15660395.28 |
| Cooling Water | Water | 0.085646 | MMGAL/H | 10.27752 | 5549860.8 |
| Steam @100PSI | Steam | 3.011514 | KLB/H | 24.513724 | 13237410.96 |
| Steam @400PSI | Steam | 5.362324 | KLB/H | 62.792814 | 33908119.56 |
| | | | | **Total** | **Rs. 6.84 crore** |

c.  **Labour**

We will assign the labours on the contract basis, as the plant will be operating only for 300 days in a year. Around 40 labourers distributed in three shifts will be required to operate the plant 24hrs a day. Total 48 lakhs will be spent on their wages annually.

| **Manpower Requirement** | **Number** | **Salary (INR)** | **Total yearly Amount** |
|---|---|---|---|
| Labour (Contract Based) | 40 | Rs 400/day | Rs 4800000 |

d.  **Maintenance**

Chemical plant require regular maintenance to operate efficiently and prevent wear and tear of equipment. A total of 4.72 Crs. will be spent annually in maintaining the plant from the external resources.

| Maintenance (External Agency) | Yearly | Rs. 4.72 crore |
|---|---|---|

e.  **Insurance**

Considering Insurance amount to be 1% of the fixed cost, which covers the insurance of the land, working staff and equipments. The insurance amount is Rs. 1.83 crores yearly.

**10. REVENUE**

| **Product Price (INR)** |
|---|

|  | kmol/hr | Quantity (kg per ann) | Market price (₹ per kg) | Selling price (₹) |
|---|---|---|---|---|
| Para-Aminophenol | 22.707 | 17820453.6 | 230 | 4098704328 |
| Aniline | 9.558 | 6400036.8 | 40 | 256001472 |
|  |  |  | Net Revenue(per year) | Rs 435.47 crore |

The aniline imported from China till 2020 was priced at nearly Rs 260 per kg and amidst the COVID-19 pandemic, its cost has increased to Rs 560 per kg. But in this project, we have kept our selling price of Para-Aminophenol as Rs 230 per kg, which is a competitive price. Moreover, we are also generating additional revenue by selling the side product Aniline generated in the reactor.

Thus, we are generating an annual revenue of **Rs 435.47 crores**.

## 11. PROFITABILITY ANALYSIS

| Profitability Analysis (INR and Crore) | | | | | | | |
|---|---|---|---|---|---|---|---|
| Year | Annual Gross | Depreciation | Depreciation Amount | Taxable Income | Taxes paid | Operating | Fixed | Cash flow |
| 0 | 0 | 0 | 0 | 0 | 0 |  | 189 | -189 |
| 1 | 180.95 | 20 | 37.800 | 143.15 | 0 | 251.59 | 3 | 180.95 |
| 2 | 180.95 | 32 | 60.480 | 120.47 | 50.1025 | 251.59 | 3 | 130.8475 |
| 3 | 180.95 | 19.2 | 36.288 | 144.66 | 42.1645 | 251.59 | 3 | 138.7855 |
| 4 | 180.95 | 11.52 | 21.773 | 159.17 | 50.6317 | 251.59 | 3 | 130.3183 |
| 5 | 180.95 | 11.52 | 21.773 | 159.17 | 55.71202 | 251.59 | 3 | 125.23798 |
| 6 | 180.95 | 5.76 | 10.886 | 170.06 | 55.71202 | 251.59 | 3 | 125.23798 |
| 7 | 180.95 | 0 | 0 | 181 | 59.52226 | 251.59 | 3 | 121.42774 |
| 8 | 180.95 | 0 | 0 | 181 | 63.3325 | 251.59 | 3 | 117.6175 |
| 9 | 180.95 | 0 | 0 | 181 | 63.3325 | 251.59 | 3 | 117.6175 |
| 10 | 180.95 | 0 | 0 | 181 | 63.3325 | 251.59 | 3 | 117.6175 |

**PAYBACK PERIOD**
Payback period is the total amount of the time required to get back the cost of equipment. It can be also defined as the amount of the time required to reach break-even point. Break-Even point is when the total cumulative cash flow becomes zero.

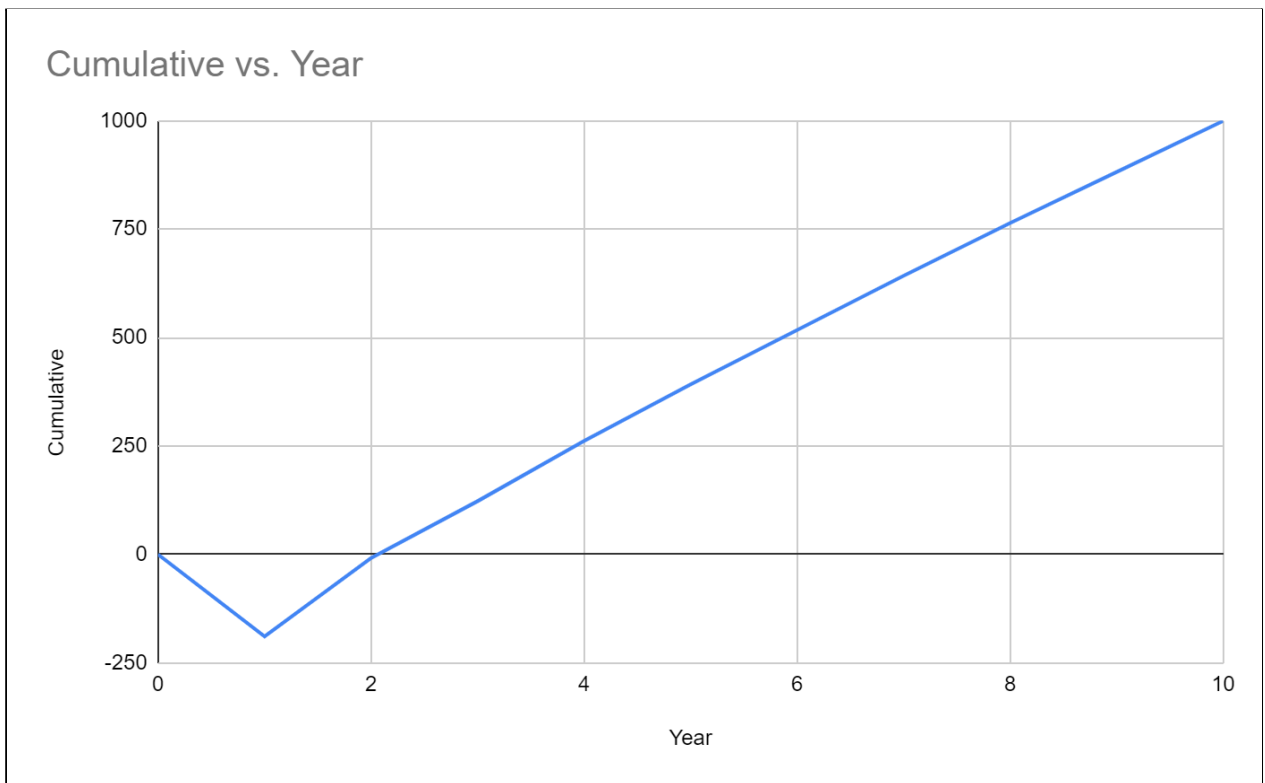

According to the above graph of Cumulative cash flow vs Year, the breakeven point is about 2. 1 years i.e. the payback time of the investment for the plant is about 2 years.

**RETURN ON INVESTMENT**
Return of investment is defined as the ratio of net income upon the investment.

$$ROI = \frac{Net\ income}{total\ investment}$$

Considering the tenure of 10 years, cumulative cash flow at the end of tenure is Rs.1116.65 crore and initial investment is about Rs. 191 Crore as fixed cost and 251.59 crore yearly as the operation cost. Hence the net investment is Rs. 2734 Crores and net income is Rs. 3851.65 crore
Hence **ROI after 10 years is 140.83%.**

## 12. SAFETY AND ENVIRONMENT

### 12.1 Nitrobenzene

Nitrobenzene needs to be kept away from the hot surface. It will be flammable if the temperature goes above the flash temperature. It must be stored in a dry and well ventilated closed container. Avoid contact with the skin and eyes and inhalation of the vapour. The area must be a non-smoking zone to prevent any hazard to the plant. Flame can be extinguished using cold water or dry carbon dioxide. [14]

### 12.2 Hydrogen

Hydrogen in a non-toxic gas. The important precaution is to be kept in a sealed container to avoid any leakages. Leakage may lead to the formation of the flammable mixture with air. Leakage can be due to process failure, container material or not proper maintenance.[15]

### 12.3 Aniline

Aniline is a flammable substance and not safe for disposal in water bodies. Heating of the chemical produces very harmful chemical vapours. Must be stored in the ventilated rooms in steel tanks. Storing temperature must be around 25-30 degree celsius. [16]

## 12.4 Para- Aminophenol

Para- AmnioPhenol needs to be carefully used, exposure to the chemical will lead to irritation in the skin and eyes, and may also lead to asthma. Hence, the people operating need to wear protective clothing. The storage container needs to be kept away from the heat source.

## 12.5 Environmental Considerations

The purged out streams contain some impure amount of nitrobenzene and Hydrogen. Nitrobenzene cannot be directly disposed of into the environment. The purged out hydrogen and
nitrobenzene can be further processed to make it more pure, and can be used in further consumption. In our preliminary design, we have proposed that hydrogen production is done using steam-methane reforming. The steam-methane reforming produces significant amounts of greenhouse gas of carbon monoxide ,carbon dioxide. The regulations on the amount of greenhouse gas emissions will become a limiting factor for the production of hydrogen. While designing the process environmental impacts need to be minimized while keeping the energy requirements as low as possible. A more sustainable method to produce hydrogen would be by electrolysis of water, which does not involve any emission of greenhouse gases.

## 13. ADDITIONAL POINTS

In order to support the costs involved with setting up this plant having a fixed capital cost of Rs 183 crores annually, we can take loan from the bank. Let us consider a loan amounting to Rs 170 crores has been approved for this project at an interest rate of 9%. Let's consider that we wish to clear this loan in 5 years. Hence with appropriate calculations, we get the following :

| Bank Loan | |
|---|---|
| Amount | Rs 170 crores |
| Tenure | 5 years |
| Rate | 9% per annum |
| EMI | Rs 3.53 crore |
| Yearly payment as per EMI | Rs 42.36 crore |
| Total Interest Payable | Rs 42 crores |
| **Total Amount (Principal + Interest)** | Rs 212 crores |

## 14. CONCLUSIONS AND RECOMMENDATIONS

With the growing uncertainty in the relations between India and China amid the border tensions and the COVID-19 pandemic, it is important for our country to be self-reliant in terms of producing the basic raw materials required for manufacturing of drugs. After all, a robust and affordable healthcare system is key to a nation's overall well being and progress.

Thus, it is possible to have cost leadership along with a novel manufacturing process for Para-Aminophenol. We believe that developing such processes for other Key Starting Materials (KSMs) and Active Pharmaceutical Ingredients (APIs) would lead our country towards the path of self reliance in healthcare as well as economic progress. The Production Linked Incentive Scheme by the India Government will support such initiatives. Moreover, with time we can extend this to synthesis of key industrial chemicals as well.

We have tried to be very realistic while choosing the values for different parameters, quantities required and the cost of materials. i.e to say our data is backed scientific literature and practical analysis. Yet there are chances that our values might not match with reality, hence we recommend proper lab trials, testing and economic analysis to be done before setting up a plant.